\documentstyle[12pt,amssymbols]{article}
\def\ltap{\raisebox{-.4ex}{\rlap{$\sim$}} \raisebox{.4ex}{$<$}}

\def\journal{\topmargin .3in    \oddsidemargin .5in
        \headheight 0pt \headsep 0pt
        \textwidth 5.625in 
\textheight 8.25in 
        \marginparwidth 1.5in
        \parindent 2em
        \parskip .5ex plus .1ex         \jot = 1.5ex}
\journal

\newskip\humongous \humongous=0pt plus 1000pt minus 1000pt

\newif\ifdtup

\def\ra{\rightarrow}
\begin{document}
\begin{titlepage}
\begin{center}
August 6, 1996       \hfill    LBNL- 39204\\

\hfill hep-ph/9608324
\vskip .5in

{\large \bf Gauge invariant formulation of strong $WW$ scattering}
\footnote
{This work was supported by the Director, Office of Energy
Research, Office of High Energy and Nuclear Physics, Division of High
Energy Physics of the U.S. Department of Energy under Contract
DE-AC03-76SF00098.}

\vskip .5in

Michael S. Chanowitz\footnote{Email: chanowitz@lbl.gov}

\vskip .2in

{\em Theoretical Physics Group\\
     Ernest Orlando Lawrence Berkeley National Laboratory\\
     University of California\\
     Berkeley, California 94720}
\end{center}

\vskip .25in

\begin{abstract}
Models of strong $WW$ scattering in the $s$-wave can be represented
in a gauge invariant fashion by defining an effective scalar propagator that 
represents the strong scattering dynamics. 
The $\sigma(qq \ra qqWW)$ signal may then be computed in U-gauge from the 
complete set of tree amplitudes, just as in the standard model, without using
the effective $W$ approximation (EWA). The U-gauge ``transcription'' has 
a wider domain of validity than the EWA, and it provides complete 
distributions for the final state quanta,
including experimentally important jet distributions that cannot be obtained
from the EWA.
Starting from the usual formulation in terms of unphysical Goldstone boson 
scattering amplitudes, the U-gauge transcription is verified by using 
BRS invariance to construct the complete set of gauge 
and Goldstone boson amplitudes in $R_{\xi}$ gauge.
\end{abstract}

\vskip .2in

\end{titlepage}

\renewcommand{\thepage}{\roman{page}}
\setcounter{page}{2}
\mbox{ }

\vskip 1in

\begin{center}
{\bf Disclaimer}
\end{center}

\vskip .2in

\begin{scriptsize}
\begin{quotation}
This document was prepared as an account of work sponsored by the United
States Government. While this document is believed to contain correct
 information, neither the United States Government nor any agency
thereof, nor The Regents of the University of California, nor any of their
employees, makes any warranty, express or implied, or assumes any legal
liability or responsibility for the accuracy, completeness, or usefulness
of any information, apparatus, product, or process disclosed, or represents
that its use would not infringe privately owned rights.  Reference herein
to any specific commercial products process, or service by its trade name,
trademark, manufacturer, or otherwise, does not necessarily constitute or
imply its endorsement, recommendation, or favoring by the United States
Government or any agency thereof, or The Regents of the University of
California.  The views and opinions of authors expressed herein do not
necessarily state or reflect those of the United States Government or any
agency thereof, or The Regents of the University of California.
\end{quotation}
\end{scriptsize}

\vskip 2in

\begin{center}
\begin{small}
{\it Lawrence Berkeley National Laboratory is an equal opportunity employer.}
\end{small}
\end{center}

\newpage

\renewcommand{\thepage}{\arabic{page}}
\setcounter{page}{1}

\noindent {\it \underline {Introduction} }

The traditional approach to strong $WW$ scattering begins by assuming 
a model for the scattering of the corresponding unphysical Goldstone bosons,
and then uses the equivalence theorem (ET) and the effective $W$ approximation
(EWA) to compute the cross section for strong production of the longitudinally
polarized gauge bosons, $\sigma(qq \rightarrow qqW_LW_L).$\cite{mc-mkg2} 
In this paper I
present a gauge invariant formulation for $s$-wave strong $W_LW_L$ 
scattering amplitudes ($L$ denotes longitudinal polarization) that allows the 
$\sigma(qq \rightarrow qqW_LW_L)$ signal to be computed directly from
the complete set of tree amplitudes without using the EWA. The strong dynamics
is carried by an effective scalar propagator, and the complete $qq \ra qqWW$
tree amplitude can be computed in any covariant $R_{\xi}$ gauge or in unitary 
gauge. This formulation is more accurate and 
provides more information than the traditional method using the EWA.

Both approaches begin with a model Goldstone boson scattering amplitude,
${\cal M}^X_R(ww \ra ww)$, required to obey unitarity
and the low energy theorems of chiral symmetry. ($X$ labels the model, $R$
denotes a covariant renormalizable gauge, most appropriately Landau gauge, and
$w_i$ is the unphysical Goldstone boson corresponding to gauge boson $W_{iL}$.) 
The equivalence theorem\cite{ET} asserts the equality of ${\cal M}^X_R(ww)$ at
high energy to the corresponding amplitude of longitudinally polarized gauge
bosons $W_L$,
$$
{\cal M}^X (W_L(p_1) W_L(p_2) \ldots ) = 
{\cal M}_R^X (w(p_1) w(p_2) \ldots )_R
+{\rm O}\left({m_W\over E_i}\right).\eqno(1)
$$
In the traditional approach the subprocess cross section 
$\sigma(W_LW_L \rightarrow W_LW_L)$ is convoluted with the effective $W_LW_L$
luminosity\cite{ewa} (which is a function of $z=s_{WW}/s_{qq}$) to obtain 
the cross section for the $WW$ fusion subprocess,
$$
\sigma(qq \rightarrow qqW_LW_L) = \int dz {d{\cal L}\over dz} 
\sigma(W_LW_L \rightarrow W_LW_L).\eqno(2)
$$

The traditional method is simple and effective but it neglects 
the transverse momentum of the final state $q$ jets and the $WW$ diboson.
Knowledge of these transverse momentum distributions is important
experimentally, e.g., to determine the efficiency of jet tag and veto detection
strategies. In Higgs boson models the $p_T$ distributions 
are readily obtained by computing the complete 
set of tree diagrams for $qq \ra qqWW$, thus avoiding the EWA. 
It would be useful and interesting to compute strong
$WW$ scattering cross sections in the same way.
A U-gauge ``transcription'' to accomplish this was presented previously and 
verified by explicit computation for specific examples\cite{noewa1}. Here an 
algorithm is presented for $s$-wave scattering models to construct the complete 
family of gauge and Goldstone boson amplitudes in R$_{\xi}$ gauge 
($WWWW,\ WWWw,\ WWww,\ Wwww$) which follow
from the initial Goldstone boson model amplitude 
${\cal M}^X_R(ww\ra ww)$ by BRS invariance.
The construction allows the gauge boson amplitude to be evaluated in any
R$_{\xi}$ gauge and in particular validates the U-gauge transcription.

In addition to providing information about the final state that is lost in the
EWA, the new method is also more accurate since it correctly sums the Higgs 
sector signal and gauge sector background amplitudes coherently, while the EWA
neglects the interference terms.\footnote{The interference term 
is important when the 
gauge sector background is large, for instance, near Coulomb singularities.}
The transcription has other interesting consequences 
that will be discussed elsewhere: it
reveals the ``K-matrix'' model, an {\it ad hoc} construction borrowed from
nuclear physics to implement partial wave unitarity and chiral symmetry, 
as a (very!) nonstandard Higgs
boson model, and allows a direct estimate of the 
effect of strong $WW$ scattering on low energy radiative corrections.

\noindent {\it \underline {The U-gauge transcription}}

The equivalence theorem, equation 
(1), already relates the gauge and Goldstone boson amplitudes,
but not in a way that can be used to extract the
$WW$ fusion amplitude ${\cal M}(qq \rightarrow qqWW)$ for strong $WW$ 
scattering. The first step is to observe that to leading order 
in the $SU(2)_L$ coupling $g$ the on-shell $WW$ amplitude is the sum of 
gauge sector and Higgs sector terms, e.g., in U-gauge 
$$
{\cal M}^X_{\rm Total}(WW\ra WW) = 
{\cal M}_{U,{\rm Gauge}}(WW\ra WW) + {\cal M}^X_{U,{\rm H}}(WW\ra WW).\eqno(3)
$$
and that at high energy, $E \gg m_W$, the gauge sector amplitude 
for longitudinal modes is dominated by its ``bad high energy behavior'',
a term growing like $E^2$ which is 
at the same  time the low energy theorem 
amplitude ${\cal M}_{\rm LET}$ of a strongly coupled 
Higgs sector\cite{mc-hg-mg,mc-mkg2}. Using the equivalence theorem 
the U-gauge Higgs sector amplitude for model $X$ is then 
$$
{\cal M}^X_{U,{\rm H}}(W_LW_L\ra W_LW_L) = {\cal M}^X_R(ww\ra ww) - 
{\cal M}_{\rm LET} + {\rm O}(g^2,{m_W \over E}),
\eqno(4)
$$
which is just the Goldstone boson model 
amplitude ${\cal M}^X_R(ww)$ with its leading threshold behavior subtracted.

This is still not in a form that can be readily embedded in
${\cal M}(qq \rightarrow qqWW)$. To proceed we limit the discussion to 
$s$-wave amplitudes and define
an effective ``Higgs'' propagator $P_X(s)$ by using the standard 
model Higgs sector amplitude for $W^+W^- \ra ZZ$ as a ``template'' for the
effective theory, 
$$
{\cal M}^X_{U,{\rm H}}(W_LW_L\ra W_LW_L) = -g^2 m_W^2 \epsilon_1 \cdot 
                          \epsilon_2 \  \epsilon_3 \cdot \epsilon_4 \ P_X
\eqno(5)
$$
where 1,2 denote the initial state bosons and 3,4 the final state. Here  
$W_LW_L \ra W_LW_L$ represents generically the two channels with $s$-wave 
threshold behavior,\footnote
{Within these channels the restriction to $s$-wave models is not very onerous,
since in these channels the LHC and electron colliders with energy $\ltap 3$
TeV will only probe energies for which the $ff \ra ffWW$ signals are dominantly
in the $WW$ $s$-wave.} 
$W^+_LW^-_L \ra Z_LZ_L$ and $W^+_LW^+_L \ra W^+_LW^+_L$.
For $W^+W^+$ this is a big departure from the standard model since the
$s$-channel exchange carries electric charge $Q=+2$, an effective ``$H^{++}$''
exchange. 

For simplicity I assume 
the weak gauge group is just $SU(2)_L$ so that $m_W=m_Z=gv/2$.
(I have verified that the conclusions do not depend on this assumption.)
Then for $E \gg m_W$ and up to
corrections of order $g^2$ the effective propagator is 
$$
P_X(s) = - {v^2 \over s^2}({\cal M}^X_R(ww\ra ww)
                 - {\cal M}_{\rm LET}).
\eqno(6)
$$

The low energy theorem\cite{mc-hg-mg,mc-mkg2}  amplitudes are 
$$
{\cal M}_{\rm LET} = \eta \ {s \over v^2} 
\eqno(7)
$$
where $\eta = +1$ for $W^+W^- \ra ZZ$ and $\eta = -1$ for $W^+W^+ \ra W^+W^+$.
Notice that ${\cal M}_{\rm LET}$ contributes $\pm 1/s$ to 
$P_X$, corresponding to a massless scalar pole, making explicit the
connection between the spontaneously broken symmetry that implies 
${\cal M}_{\rm LET}$ and the cancellation of the bad high energy behavior 
by Higgs boson exchange. The residual contribution to $P_X$ from
${\cal M}^X_R$ carries the model dependent strong interaction dynamics.

With the effective propagator $P_X$ we can formulate the U-gauge transcription 
in a way that allows us to embed ${\cal M}^X_{U,{\rm H}}(W_LW_L)$ into 
${\cal M}(qq \ra qqWW)$. The prescription is simple: compute the usual 
gauge sector tree diagrams for ${\cal M}(qq \ra qqWW)$ in U-gauge 
but replace the
Higgs boson exchange diagram(s) by $s$-channel exchange of $P_X$ with the 
$WW$``$H$'' vertex given by $gm_W\ g^{\mu\nu}$ as in equation (5).
Here we make the usual, unavoidable  extrapolation, required in any approach 
to strong $WW$ scattering, from massless (in Landau gauge) Goldstone boson 
to $m_W$ in the on-shell gauge boson amplitude 
to space-like $-q^2 \simeq O(m_W^2)$ for the initial 
state virtual $W$'s in the $WW$ fusion amplitude. As always, the 
extrapolation contributes to the inevitable $O(m_W/E)$ correction.

\noindent {\it \underline {BRS and gauge invariance}}

In \cite{noewa1} the U-gauge transcription 
was verified by explicit calculation for the 
K-matrix model and the heavy Higgs boson standard model in the $W^+W^+$ channel.
We now demonstrate its validity in general for $s$-wave scattering 
by constructing a 
BRS invariant\cite{brs} set of amplitudes that relate the input model 
${\cal M}^X_R(ww\ra ww)$ to the corresponding gauge boson amplitude 
${\cal M}^X(W_LW_L \ra W_LW_L)$. Beginning from the Goldstone boson amplitude
${\cal M}^X_R(ww\ra ww)=$ $<wwww>$ we use BRS invariance 
to construct the family of amplitudes
$<wwwW>,\ <wwWW>,\ <wWWW>$ and $<WWWW>$ in the generalized $R_{\xi}$ gauge.
The gauge boson amplitude ${\cal M}^X(WW\ra WW)$ is then explicitly 
gauge invariant ($\xi$ independent) and for the longitudinal modes is
precisely the previously formulated U-gauge transcription. BRS invariance is
verified explicitly for the {\em amplitudes}, even though there may 
($W^+W^- \ra ZZ$) or may not ($W^+W^+ \ra W^+W^+$) be an 
underlying effective Lagrangian.

The construction of the BRS invariant set of amplitudes is accomplished by
following a Feynman diagram algorithm, using as a template the set of diagrams 
for $W^+W^- \ra ZZ$ in the standard model. That is, we write each amplitude
($<wwww>,\ <wwwW>,\ <wwWW>,\ <wWWW>$ and $<WWWW>$) as a sum of terms
corresponding to the tree Feynman diagrams for the $W^+W^- \ra ZZ$ 
channel in the standard model. By maintaining the diagrammatic 
form of the amplitude and the
essential relationships between vertices and propagators as they are in the
standard model, we automatically preserve BRS invariance. This
$W^+W^- \ra ZZ$ standard model template is applied to strong scattering in the 
$W^+W^- \ra ZZ$ channel and, less obviously, also to the  $W^+W^+\ra W^+W^+$ 
channel.

The diagrammatic algorithm is specified by the Feynman rules for vertices and
propagators. The $WWH$ vertex\footnote{
$WWH$ denotes generically $W^+W^-H$ and $ZZH$ with reference to the 
$W^+W^- \ra ZZ$ channel and $W^+W^+H^{++}$ for $W^+W^+ \ra W^+W^+$.} 
and the propagator $P_X(s)$ were defined already in the U-gauge transcription,
equations 5-6. 
The three and four gauge boson vertices and the propagators of the
gauge and unphysical Goldstone bosons are determined by gauge sector dynamics
and keep their standard model values. 

The quartic Higgs sector coupling $\lambda_X$ 
and the Goldstone-Higgs $wwH$ vertex are 
related as in the standard model, $\lambda_{wwH}= -2\lambda_X v$.
The quartic coupling $\lambda_X$ is then 
determined by expressing the input model ${\cal M}^X_R(ww\ra ww)$ 
as the sum of the
four-point contact interaction and $s$-channel Higgs exchange amplitude, 
$$
{\cal M}^X_R(ww\ra ww)= -2\lambda_X\eta -(2\lambda_Xv)^2P_X  \eqno(8)
$$
where $\eta$ is defined in equation 7. Using equations (6) and (7) we solve
equation (8) to obtain 
$$
\lambda_X= {s\over 2v^2}{{\cal M}^X_R \over {\cal M}^X_R - {\cal M}_{\rm LET}}
\eqno(9)
$$
and 
$$
P_X= {\eta \over s-2\lambda_Xv^2}.
\eqno(10)
$$
While the effective coupling ``constant'' $\lambda_X$ is not in fact constant
but is in general a function of $s$, equation 10
shows that we have preserved (up to the factor $\eta$) the 
standard model relationship between the effective Higgs propagator and 
the Higgs sector vertices which is crucial for maintaining BRS invariance.

The remaining interaction vertices are fixed by requiring that the 
$WWWW$, $WWWw$, $WWww$, $Wwww$, and $wwww$ amplitudes satisfy 
the BRS identities
$$
(\partial W + \xi m_W w)^n=0 \eqno(11)
$$
for n = 1,2,3.4. 
For the $W^+W^- \ra ZZ$ channel all vertices not specified above are given
precisely by their standard model values. 
The amplitudes obtained from our algorithm then trivially satisfy BRS
invariance and ${\cal M}(W^+W^- \ra ZZ)$ is trivially gauge invariant 
($\xi$ independent in $R_{\xi}$ gauge), because equation 10
assures that the  necessary 
cancellations occur just as in the standard model.

It is less trivial but no less straightforward to verify that the prescription
can be made to work for $W^+W^+ \ra W^+W^+$. In this case it is necessary to
define some nonstandard interaction vertices, since the U-gauge
transcription defined above already mutilates the standard model structure for 
$W^+W^+ \ra W^+W^+$ by substituting
an $s$-channel effective $H^{++}$ exchange for the $t$ and $u$-channel 
$H^0$ exchanges
of the standard model. Clearly all interactions of the effective $H^{++}$ boson 
are nonstandard and require definition. The U-gauge transcription
already specifies the $H^{++}W^-W^-$ vertex as $gm_Wg^{\mu\nu}$ 
(see equation 5), equal\footnote{I follow the phase conventions of the CORE
compendium.\cite{core}} to the standard model $HW^+W^-$ vertex. The remaining
nonstandard vertices are fixed by insisting on the validity of the 
BRS identities, equation 11, applied to $<W^+W^+W^-W^->$ for n = 1,2,3,4.
With the vertices chosen to satisfy BRS invariance we find that $\xi$
independence of ${\cal M}(W^+W^+ \ra W^+W^+)$ in $R_{\xi}$ gauge is also
automatically assured.

In addition to defining the interactions of the effective $H^{++}$ boson
we must adopt nonstandard quartic couplings
for the $w^+w^+w^-w^-$ and $W^+W^+w^-w^-$ vertices: the former is -1/2 times
its standard model value while the latter does not exist 
at all in the standard model. Vertices that do not exist in the standard 
model or that differ from their standard model values are given in 
table~1. 

To illustrate how the diagrammatic  algorithm satisfies BRS invariance, 
consider the identity equation 11 with $n=2$, applied to the two initial 
state bosons in $WW$ scattering. We can consider
$W^+W^- \ra ZZ$ and $W^+W^+ \ra W^+W^+$ concurrently, since 
in our approach they are given by the same set of Feynman diagrams. 
The BRS identity for the scattering amplitude $W_1W_2 \ra W_3W_4$ is 
$$
\epsilon_{3\alpha}\epsilon_{4\beta}\left(
k_{1\mu}k_{2\nu}{\cal M}^{\mu\nu\alpha\beta} 
                 +im_W(k_{1\mu}{\cal M}^{\mu\alpha\beta}_{w_2}
                       k_{2\nu}{\cal M}^{\nu\alpha\beta}_{w_1})
                  -m_W^2{\cal M}^{\alpha\beta}_{w_1 w_2}\right) = 0.
\eqno(12)
$$
The subscript $w_i$ indicates the amplitude in which gauge boson $W_i$ is
replaced by Goldstone boson $w_i$.

Using the Feynman rules defined above to evaluate the amplitudes in 
equation 12 in $R_{\xi}$ gauge, we find after trivial cancellations that 
the remaining terms are 
$$
\delta_{\rm BRS}^2 = {1\over 2}g^2m_W^2\epsilon_3 \cdot \epsilon_4 \left(
                       (s - 2\lambda_X v^2)P_X - \eta \right). \eqno(13)
$$
Using equation 10 the right side vanishes, confirming the BRS
identity equation 12. All other BRS identities can be similarly verified.

\newpage
\noindent {\it \underline {Discussion}}

The U-gauge transcription has been verified for $s$-wave strong $WW$ 
scattering models by demonstrating its consistency with BRS invariance for
the complete set of gauge and Goldstone boson scattering amplitudes.
A Feynman diagram algorithm was defined, including an effective
``Higgs'' propagator that carries the strong scattering dynamics and a related
energy dependent ``effective'' $\phi^4$ Higgs sector coupling constant.
The transcription is useful both in high energy applications, to strong $WW$
scattering at $pp$ and $e^+e^-$ colliders, and in low energy applications, 
to estimate the effect of strong $WW$ scattering on electroweak radiative
corrections. 

As discussed in \cite{noewa1} the U-gauge transcription is more accurate and
more complete than the effective $W$ approximation for computing strong $WW$
scattering. It is more accurate because it retains the interference between the
strong $WW$ scattering amplitude and the gauge sector background amplitude, 
and also because it provides the transverse momentum of the $WW$ diboson which
is neglected in the EWA. It is more complete because it provides the full
three-momentum distribution for the final state quark jets in the $qq \ra qqWW$
process, while the EWA neglects the jet 
transverse momenta (also introducing an error in the determination of the jet
rapidities). 

The final state jet distributions are needed to 
compute the efficiency of detection strategies such as the central jet
veto\cite{cjv} and the forward jet tag\cite{tag}. The former is very effective
against the gluon exchange and electroweak gauge sector backgrounds, and the
latter may be very useful against the surprisingly large $\overline qq \ra WZ$
background\cite{wz} to the $W^+W^+ \ra W^+W^+$ strong
scattering signal.
In previous studies the necessary jet distributions have been estimated assuming
the same shape for strong scattering as for the standard model with a heavy 
(typically 1 TeV) Higgs boson. This assumption can now be tested using the
U-gauge transcription. I find that it works well at low enough energy colliders,
for which the strong scattering and heavy Higgs cross sections are ``squashed''
into roughly the same region in $s_{WW}$, but not at higher energy colliders 
with enough phase space to allow the differences in the $WW$ energy spectrums
to emerge. From this perspective the LHC is a ``low'' energy collider, 
while at SSC energies (R.I.P.) the differences begin to be important.

The effective Higgs sector propagator defined in the U-gauge transcription may
also be used to estimate the direct effect of 
strong $WW$ scattering on low energy
radiative corrections. Unlike the typically large corrections predicted by
technicolor models, the correction due just to strong nonresonant dynamics in 
$WW$ scattering is  not very much bigger than the effect of the 1 TeV standard
model Higgs boson. These results and other applications of the U-gauge
transcription will be presented elsewhere.

\vskip .2in
\noindent Acknowledgements: I wish to thank R.N. Cahn M.K. Gaillard, and H. 
Murayama for helpful discussions and J. Ellis for perusing the manuscript. 
This work was supported by the Director, Office of Energy
Research, Office of High Energy and Nuclear Physics, Division of High
Energy Physics of the U.S. Department of Energy under Contracts
DE-AC03-76SF00098 and DE-AC02-76CHO3000.

\newpage

Table 1. Nonstandard interaction vertices for $W^+W^+ \ra W^+W^+$ scattering.
Also shown are analogous vertices for $W^+W^- \ra ZZ$, which agree precisely 
with the standard model. All momenta are 
inflowing, phase conventions are as in the CORE compendium\cite{core}, and 
$\eta$ is defined below equation 7.

\begin{center}
\vskip 20pt
\begin{tabular}{ccc}
$W^+W^+ \ra W^+W^+$ & $W^+W^- \ra ZZ$ & Interaction\cr
&&\cr
\hline
\hline
&&\cr
$H^{++}(k)W^-_{\mu}(p)W^-_{\nu}(q)$ & $H^0(k)W^+_{\mu}(p)W^-_{\nu}(q)$ & 
 $ gm_Wg^{\mu\nu}$ \cr
&&\cr
\hline
&&\cr
$H^{++}(k)W^-_{\mu}(p)w^-(q)$ & $H^0(k)W^+_{\mu}(p)w^-(q)$ & 
 $ i{g }(q^{\mu} - k^{\mu})/2$ \cr
&&\cr
\hline
&&\cr
$H^{++}(k)w^-(p)w^-(q)$ & $H^0(k)w^+(p)w^-(q)$ & $-2\lambda_Xv$ \cr
&&\cr
\hline
&&\cr
$w^+(p_1)w^+(p_2)W^-_{\mu}(p_3)W^-_{\nu}(p_4)$ & 
 $w^+(p_1)w^-(p_2)Z_{\mu}(p_3)Z_{\nu}(p_4)$ &
 ${\eta } g^2 g^{\mu\nu}/2$ \cr
&&\cr
\hline
&&\cr
$w^+(p_1)w^+(p_2)w^-(p_3)w^-(p_4)$ & $w^+(p_1)w^-(p_2)z(p_3)z(p_4)$ &
$ -2\eta\lambda_X$ \cr
&&\cr
\hline
\hline 
  
\end{tabular}
\end{center}

\end {document}